\documentclass[sigconf]{acmart}
\usepackage[utf8]{inputenc}
\usepackage[english]{babel}
\usepackage[T1]{fontenc}

\usepackage{tablefootnote}
\usepackage{multirow}
\usepackage{booktabs}
\usepackage{adjustbox}
\usepackage{graphicx}
\usepackage{csquotes}

\newtheorem{definition}{Definition}

\usepackage{xcolor}

\usepackage[inkscapelatex=false]{svg}
\usepackage{subcaption}
\usepackage{todonotes}

\usepackage{glossaries}
\makeglossaries

\newglossaryentry{Tuple-DP}
{
	name=Tuple-DP,
	description={Mécanisme DP qui considère une entrée (une ligne, dans une table) comme l'unité de confidentialité}
}

\newglossaryentry{User-DP}
{
	name=User-DP,
	description={Mécanisme DP qui considère une personne comme unité de confidentialité. Ainsi deux base sont voisine, si l'on retire l'ensemble des données associé directement ou indirectememnt à une données identifiantes (PID).}
}
\newglossaryentry{proxy}
{
	name=Proxy,
	description={Architecture de système, se positionnant entre l'utilisateurice, et le \acrshort{SGBD}. Un proxy peut modifier
	la requête SQL, nettoyer les résultat, etc ... mais n'execute pas lui même la requête}
}

\newglossaryentry{donnee-pub}
{
	name=Données publiable,
	description={En confidentialité différentielle, il est compliqué (et peut-être non souhaitable ?) d'apporter
	une notion de données publiques et privées \\
	Il me manque actuellement une bonne définition de données publiable \dots 
	}
}

\newacronym{SGBD}{SGBD}{Système de Gestion de Base de Données}
\newacronym{DBMS}{DBMS}{Data Bases Management System}

\newacronym{GS}{GS}{Global Sensitivity}
\newacronym{LS}{LS}{Local Sensitivity}
\newacronym{SS}{SS}{Smooth Sensitivity}
\newacronym{SAA}{SAA}{Sample And Aggregate : a framework to compute differential private aggregation fonction
\cite{NissimRS07}}

\newacronym{SA}{SA}{Selection, Aggregation}
\newacronym{SJA}{SJA}{Selection, Jointure, Aggregation}
\newacronym{SPJA}{SPJA}{Selection, Projection, Jointure, Aggregation}

\newacronym{LP}{LP}{Programme Linéaire}

\newcommand{\sql}[1]{\verb|#1|}

\begin{document}
\title{Experiments \& Analysis of  Privacy-Preserving SQL Query Sanitization Systems}
\author[L. Ecoffet]{Loïs Ecoffet}
\orcid{0009-0006-8436-9236}
\email{lois.ecoffet@umlp.fr}
\author[V. Rehn-Sonigo]{Veronika Rehn-Sonigo}
\email{veronika.sonigo@umlp.fr}
\author[J.-F. Couchot]{Jean-François Couchot}
\orcid{0000-0001-6437-5598}
\email{jean-francois.couchot@umlp.fr}
\affiliation{%
  \institution{Université Marie et Louis Pasteur, CNRS, institut FEMTO-ST (UMR 6174)}
  \city{Besançon}
  \country{France}
}
\author[C. Palamidesi]{Catuscia Palamidessi}
\orcid{0000-0003-4597-7002}
\email{catuscia.palamidessi@polytechnique.edu}
\affiliation{%
  \institution{Inria Saclay, 1 rue Honoré d'Estienne d'Orves			
 Bâtiment Alan Turing, Campus de l'École Polytechnique}
  \city{Palaiseau}
  \country{France}
}

\acmConference{EDBT/ICDT 2026 Joint Conference}{24th March - 27th March 2026}{Tampere, Finland}
\begin{abstract}
Analytical SQL queries are essential for extracting insights from relational databases but concurrently introduce significant privacy risks by potentially exposing sensitive information. 
To mitigate these risks, numerous query sanitization systems have been developed, employing diverse approaches that create a complex landscape for both researchers and practitioners. 
These systems vary fundamentally in their design, including the underlying privacy model, such as k-anonymity or Differential Privacy; the protected privacy unit, whether at the tuple- or user-level; and the software architecture, which can be proxy-based or integrated. 
This paper provides a systematic classification of state-of-the-art SQL sanitization systems based on these qualitative criteria and the scope of queries they support. 
Furthermore, we present a quantitative analysis of leading systems, empirically measuring the trade-offs between data utility, query execution overhead, and privacy guarantees across a range of analytical queries. 
This work offers a structured overview and performance assessment intended to clarify the capabilities and limitations of current privacy-preserving database technologies.
\end{abstract}

\maketitle

\section{Introduction}

Relational databases are fundamental tools for modern data management, designed to efficiently store, retrieve, and analyze vast amounts of information. 
Analytical queries, particularly those combining Select, Project, Join, and Aggregate (SPJA) operations, are essential for extracting valuable insights for decision support, business intelligence, and reporting. 
However, the very power that makes these queries useful also introduces significant privacy risks.

While providing powerful data access, these queries can inadvertently expose sensitive information, leading to potential privacy breaches and violations of legislative mandates such as the GDPR. 
Consequently, protecting data while preserving its utility has become a critical challenge. 
In this context, raw data exports, or even DUMP operations, which capture the entire state of a database, represent a maximal privacy risk and are typically reserved for backup and recovery rather than analytical sharing.

To address this challenge, query result sanitization has emerged as a crucial technique for enabling data analysis while protecting sensitive information. 
This process involves modifying query outputs to prevent the inference of personal data, thereby balancing the competing demands of data utility and privacy.

The implementation of query sanitization is multifaceted, defined by several key design choices that shape a system's capabilities and trade-offs. 
Two prominent theoretical frameworks guide this process: 
k-anonymity~\cite{k-anonymity}, which ensures that individuals are indistinguishable within a group by clustering data, and Differential Privacy~\cite{DBLP:conf/tcc/DworkMNS06}, which provides formal, probabilistic guarantees by injecting controlled noise. 
The application of these models can be further distinguished by the privacy unit — either protecting individual records (tuple-level) or all data associated with a person (user-level). 
Finally, these systems are deployed using two primary software architectures: proxy-based systems that intercept and rewrite queries externally, offering broad compatibility, or integrated systems that are built into the database itself, allowing for deeper control over query execution.

Given the diversity of these approaches, selecting and deploying an appropriate SQL sanitization system is a complex task. A clear understanding of their functional scope, practical limitations, and performance trade-offs is essential for both researchers and practitioners. 
This paper addresses this need by providing:
\begin{itemize}
    \item \textbf{A comprehensive classification of SQL sanitization systems}: We conduct a systematic classification of state-of-the-art systems based on the following qualitative criteria: the underlying privacy model, the privacy unit (tuple or user), the software architecture, and the scope of supported SQL queries.
    \item \textbf{Empirical evaluation of performance and utility trade-offs}: We provide a quantitative analysis of leading privacy-preserving systems, empirically measuring the trade-offs between data utility, query execution overhead, and privacy guarantees across a range of analytical queries.
    \item \textbf{Identification of key limitations and gaps}: Our analysis highlights the significant gap between the theoretical promise and practical capabilities of current systems, particularly their failure to support complex SQL structures (e.g., subqueries) and their performance bottlenecks, thereby identifying critical areas for future research.
\end{itemize}

The reminder of this article is organised as follows:
Section~\ref{sec:background} establishes the foundational concepts and qualitative criteria necessary for our analysis, covering privacy models, system architectures, and supported query types. 
Based on these criteria,  Section~\ref{sec:systems} introduces and classifies several state-of-the-art SQL query sanitization systems. 
Section~\ref{sec:qualitative} then provides a comparative summary and qualitative analysis of these systems. 
In Section~\ref{sec:quantitative}, we present our quantitative evaluation, detailing the experimental methodology and assessing the systems' performances and utility trade-offs. 
Finally, Section~\ref{sec:discussion} discusses the broader implications and limitations of current approaches, before we summarize the paper in Section~\ref{sec:conclusion} and provide some future work.

\section{Background}\label{sec:background}

This section establishes the foundational concepts and qualitative criteria necessary for understanding and classifying privacy systems designed to sanitize SQL queries. To provide a focused and relevant comparative analysis (Section~\ref{sec:systems}), our study considers four key dimensions that define the landscape of privacy-preserving database systems: the Privacy Model (Section~\ref{sec:model})), the Privacy Unit (Section~\ref{sec:unit}), the System Architecture (Section~\ref{sec:archi}), and the supported Query Types (Section~\ref{sec:type}).

\subsection{Interactive vs Non-Interactive Settings}
Starting with this quote of Dwork \emph{et al.}~\cite{DworkMNS16}:
\textquote{
In the non-interactive setting, the curator—a trusted entity—publishes a “sanitized” version of the collected data; 
	the literature uses terms such as “anonymization”, “de-identification”, and “synthetic data”. 
	[...] In the interactive setting, 
	the data collector provides a mechanism with which users may pose queries about the data, and 
	receive (possibly noisy) responses.
}

Interactive and non-interactive databases require very different approaches to privacy protection.  
Non-interactive databases are notably the result of sanitized DUMPs. 
Given that our study focuses on mechanisms applied at query level, 
this document will exclusively focus on the interactive setting. 
We now proceed to detail the qualitative criteria used for system comparison.

\subsection{Privacy Model}\label{sec:model}
The privacy model aims to provide a theoretical and technical framework to measure privacy protection.
Each model brings a nuance to this definition, as well as technical possibilities or constraints.

\subsubsection{$k$-Anonymity: }
$k$-anonymity \cite{k-anonymity} provides privacy by ensuring that each individual in a dataset cannot be 
distinguished from at least $k-1$ other individuals. 

To achieve this, $k$-anonymity defines quasi-identifiers, elements that are not in themselves unique identifiers, 
but which are sufficiently well correlated with an entity that they can be combined with other quasi-identifiers 
to  create a unique identifier.
\begin{definition}[Quasi-identifiers (QID)~\cite{nguyen2020techniques}] 
  The attributes of $Q \subset {A_1,\dots,A_m}r$ are quasi-identifiers of relation $T$ if the following query returns at least one result: 
\begin{verbatim}
SELECT Q FROM T GROUP BY Q HAVING COUNT(*)=1
\end{verbatim}
  
\end{definition}
With this definition of QIDs, we can define groups of at least $k$ people with the same QIDs. 
Each person cannot be isolated from its group, and therefore cannot be distinguished from $k-1$ persons, in the queries.

\begin{definition}[$k$-Anonymity]
	A dataset $D$ is $k$-anonymous if the information relating to each person in it cannot be distinguished 
	from at least $k - 1$ individuals whose information appears in $D$. No result should be returned by:
\begin{verbatim}
SELECT Q, COUNT(*) AS C FROM T GROUP BY Q 
  HAVING C > 0 AND C < k
\end{verbatim}
        
where \verb+Q+
        and \verb+T+ respectively stand for the QID and the relation.
\end{definition}

%
%
\subsubsection{Differential Privacy:}
Differential privacy \cite{DworkR14:DP} provides a formal guarantee of indistinguishability.
This means that no one can know whether personal data is part of a database or not. 
More precisely, the statistical results between a database with, or without, a person's data (called 
\emph{neighboring databases}) are almost the same. 
The distance function $d$ represents the difference between two databases. The definition of this function modifies the notion of differential privacy. Two databases $D_1$ and $D_2$ are neighbors if $d(D_1,D_2) = 1$.

\begin{definition}[Differential Privacy~\cite{DworkR14:DP}]
	A randomized mechanism $\mathcal{M}$ is $(\epsilon, \delta)$-differential private ($(\epsilon, \delta)$-DP) if for any neighboring
	databases $D_1$, $D_2$ and a set of possible output $S$: 
	$$
	Pr[\mathcal{M}(D_1) \in S] \leq e^\epsilon \times Pr[\mathcal{M}(D_2) \in S] + \delta
	$$
\end{definition}

The $\epsilon$ value represents the amount of data we can obtain on an individual. This value is called the “privacy budget”. The smaller the budget is, the more the data are protected. Any query on a DP-protected database consumes part of the budget. Queries must therefore be answered with caution to avoid running out of budget.
The budget can be split into several parts to meet several requests or to apply several mechanisms in succession. This is called \emph{composition} and represents a complex problem.

The value $\delta$ represents the possibility that the mechanism will leak more information than the budget. 
It is therefore important to keep this probability very low~\cite{DworkR14:DP}.
The addition of this margin of error allows for more permissive mechanisms where pure DP (with $\delta = 0$) would not. 

Notice that the definition of neighboring databases ($D_1$ and $D_2$) is intrinsically linked to the Privacy Unit (Section~\ref{sec:unit}). 
If the unit is a tuple, $D_1$ and $D_2$ differ by one tuple. If the unit is a user, they differ by all tuples associated with one user.

\subsubsection{Sensitivity:}
A standard approach to respect this property is to add noise, which is parameterized by the sensitivity of the request.
Sensitivity represents the maximum impact that a contribution can have on the query result, \textit{i.e.}, the difference between the query result on the two neighboring databases.

\begin{definition}[Global Sensitivity (GS)]
	For any neighboring databases $D_1$ and $D_2$, a query $Q$ has a global sensitivity of $GS_Q$ calculated as follows: 
	$$
	GS_Q = \max_{D_1,D_2 | d(D_1,D_2) = 1}{\|Q(D_1) - Q(D_2) \|}
	$$
\end{definition}

GS can be unbounded for some queries, resulting in infinite noise and unusable sanitized data.
Another definition of sensitivity is \acrfull{LS}. 
LS is the sensitivity of a query knowing the real database $D_1$.

\begin{definition}[Local Sensitivity (LS)]
	For a real database $D_1$ and every neighboring data\-base $D_2$ of $D_1$, a query $Q$ has a local sensitivity of $LS_Q(D_1)$ calculated as:
	$$
	LS_Q(D_1) = \max_{D_2 | d(D_1,D_2) = 1} {\|Q(D_1) - Q(D_2)\|}
	$$
\end{definition}

\subsection{Privacy Unit} \label{sec:unit}
Talking about privacy means defining the privacy unit that has to be protected. 
Personal data can be represented in many different ways, and the definitions of the privacy model are impacted by the way in which they are recorded, namely \emph{tuple-level} and \emph{user-level}.

\subsubsection{Tuple-level: }
Personal data are stored in a single tuple. Each individual is the owner of an unique tuple, and if this tuple is protected, then the individual's privacy is preserved. 

\subsubsection{User-level: }
Personal data is stored in potentially multiple tables or rows. 
Each row containing personal data can be related to a PID (personal identifier).
Some systems assume that each tuple is tagged with a PID attribute that determines who owns the data~\cite{WilsonZLDSG20}.
In relational databases, foreign keys can be used to determine whether a given row is associated with a PID or not~\cite{Dong23}.

The choice of the privacy unit defines a utility-privacy trade-off in Differential Privacy. Tuple-level DP has low sensitivity, requiring less noise for higher utility, but is weak against attacks where a user owns multiple records. User-level DP offers a robust privacy guarantee by protecting all data associated with an entity. This robustness, however, often leads to a higher Global Sensitivity bound, consequently introducing more noise and reducing data utility.

\subsection{System Architecture} \label{sec:archi}

Architecture determines how the privacy system will interact with the \acrfull{DBMS}, and what control it will have over query processing.
There are basically two types of systems namely proxy-based and integrated.

\subsubsection{Proxy: }
A proxy is a system that interposes itself between the client and the database. It intercepts communications and can modify data. 
The proxy can integrate privacy by modifying SQL queries and modifying results. 
Proxies have the advantage of being able to integrate with any system and benefit from \acrshort{DBMS} optimizations, while allowing a wide variety of privacy algorithms.

\subsubsection{Integrated: }
Integrated systems implement database management or are \acrshort{DBMS} extensions.
The analyst in this case, must go through the given integrated interface.  
This makes it possible to control some query answering processes, 
but creates an interoperability problem, as the system cannot be used with any environment. 

\subsection{Query Types} \label{sec:type}
A system offers protection for a subset of statistical queries. It can refuse certain types of queries if they would compromise privacy.
SQL queries can be categorized in several ways, depending on the relational operations performed. The global scheme of a statistical query is Selection-Projection-Join-Aggregation (SPJA), which selects tuples and attributes from potentially several linked tables, then calls an aggregation function to obtain a statistic.
There are also queries that do not use an aggregation function, such as other queries returning raw data.

\subsubsection{Statistical and Data Queries: }\label{sub:sub:statistical} 
A query can return either: \textbf{(A)} statistical data, via aggregation functions (in SQL: \verb+COUNT+, \verb+SUM+,
\verb+AVG+, etc ...)
\textbf{(D)} or it can return data, via individual tuples (processed or not). 
A DUMP is the most general form of data query, where all tables, tuples and attributes are returned.

\subsubsection{Selection: }
\emph{Selection} allows  to filter tuples according to a predicate. This makes it possible to keep only certain 
results based on data linked to the tuple, and can therefore be used to isolate a person.
It is implemented by the \verb+WHERE+ keyword in SQL.

\subsubsection{Projection: }
\emph{Projection} is the relational operation that allows to keep only one set of attributes. It is equivalent to the \verb+SELECT+ keyword in SQL.

In his thesis, Dong~\cite{DongThesis} explains that projection reduces the risk of information being revealed under Differential Privacy. 
\textquote{Intuitively, a projection reduces the query answer, hence its sensitivity, so it requires  less noise.}

The same reasoning applies to $k$-anonymity. Indeed, projecting a subset of attributes can only reduce 
the number of quasi-identifiers, and therefore create larger groups of people.

To adjust privacy mechanisms as effectively as possible, we need to take into account which attributes are projected and which are not. 
A system is considered to take projection into account \textbf{(P)} if its mechanisms are parameterized by the attributes projected by the query.

\subsubsection{Join:}
\emph{Join} connects tuples from two different tables. This is implemented by the \verb+JOIN+ keyword in SQL.
By linking one tuple to another, it is possible to correlate data and thus single out an individual.  
In the case of $k$-anonymity, this complicates the management of quasi-identifiers because a tuple can 
be related to an unbounded number of other quasi-identifiers, and thus a join can quickly reduce the size of groups.

In the case of differential privacy, joins can increase the sensitivity of queries indefinitely, 
\textquote[\cite{JohnsonNS18:Flex}]{because the output of a join may contain duplicates of sensitive rows.}

However, there are several types of joins. Some systems take only a subset of them into account (notably one-to-one joins, where a tuple can only be associated with another tuple) \cite{JohnsonNS18:Flex}.
To consider that a system supports joins \textbf{(J)}, it must provide protection for general joins.  

\subsubsection{Histogram:}\label{sub:sub:histo}

A histogram divides the database into different categories and returns the number of elements for each category. 
The category key is the attribute used to group tuples in the database. 
There are two types of histograms depending on whether the domain of category keys is finite or not. 
These two types have no syntactic differences in SQL. The domain is linked to the database schema or to data modification policies.
However, there is a difference to protect them.
\begin{itemize}
	\item \textbf{(H1) finite domain: } The domain is finite, and can be known by the system when the query is executed. For example, a query that returns the histogram of numbers of visits according to days of the week, 
		a contribution cannot create a new category.
	\item \textbf{(H2) non-finite domain: } The domain is either infinite or unknown at query execution time. For example, if a query returns the histogram of favorite vegetables, 
		a contribution can create a new category.
\end{itemize}

\section{Database query sanitizing systems}\label{sec:systems}
This section introduces some SQL query sanitization systems. 
For each system, a brief presentation of its functionality will be given, with the privacy model, privacy unit, architecture and how different types of queries are protected. 
The last part provides a table summarizing the features of each system.

\subsection{\textit{k}-anonymity based systems}

$k$-anonymity-based systems take two different approaches. 

They either enforce privacy by statically grouping records with identical 
Quasi-Identifiers  into anonymization sets (e.g., PostgreSQL Anonymizer), or they dynamically control the size of the set targeted by a query, rejecting it if it targets fewer than 
$k$ individuals (e.g., Diffix, Open Diffix).

The approaches used to ensure $k$-anonymity are information suppression 
(for example, masking the telephone number) and generalization (for example, making age ranges). 

\subsubsection{Diffix}
Diffix \cite{FrancisEM17:Diffix} is a \emph{proxy}-based system designed  for \emph{user-level} protection, supporting statistical queries \textbf{(A)} on database . 

To this, it provides \emph{$k$-anonymity}
preventing queries whose number of targeted people is less than a parameter $k$.
Diffix adds a layer of protection to \emph{k-anonymity} by  \textquote[\cite{FrancisEM17:Diffix}]{[adding] noise [to statistical results], but [making] the noise dependent on the data}

The system is designed to protect all users from being isolated (classic isolation, or intersection attack). 
It supports \emph{Selection} \textbf{(S)}.
Due to its focus on single-table data, 
 \emph{general joins} are so not supported.

For histograms, it applies a  \emph{noisy threshold} to counts, suppressing categories falling below the threshold \textbf{(H2)}.

\subsubsection{Diffix updates:}
This initial version of Diffix is followed by several updates (Birch\cite{Diffix-Birch}, Cedar\cite{Diffix-Cedar}
and Dogwood\cite{Diffix-Dogwood}), 
with the aim to improve the support for complex SQL queries, including joins.
However, an in-depth analysis of the latest Diffix update would be necessary to find out whether this latest version can protect joins. The same should be done for histograms, statistical queries and projections.

\subsubsection{Open Diffix}
Open Diffix is based on Diffix Elm, which itself  is a simplification and modification of Diffix. 
The system still maintains \emph{user-level} \emph{$k$-anonymity}, augmented with sticky noise applied to aggregation results. 
Its privacy model is therefore \textquote[\cite{Diffix-Elm}]{stronger than $k$-anonymity [\ldots] but not as strong as Differential Privacy (DP)}.

Architecturally, it  is now \emph{integrated} into the \acrshort{DBMS}, in the form of a PostgreSQL extension. 
Query support is limited to statistical queries \textbf{(A)}, using count as the sole aggregation function. 
Sticky noise is proportional to the targeted user set, allowing the system to adjust protection 
based on selection \textbf{(S)} and projection \textbf{(P)}.
Joins are not supported. 

Histograms \textbf{(H2)} are protected with a sticky noisy threshold, \textit{i.e.}, categories with too few people are hidden. The threshold value is randomly drawn around a fixed value, to prevent giving information 
if a category is observed to have disappeared. 
The difference with \emph{$\tau$-threshold}~\cite{WilsonZLDSG20} is that the threshold is noisy, not the value.

\subsubsection{PostgreSQL Anonymizer}

PostgreSQL Anonymizer \cite{PostgreSQL_Anon} is a PostgreSQL extension (\emph{integrated} architecture) that uses masking and generalization 
to implement \emph{$k$-anonymity} at \emph{tuple-level}. 
The system provides privacy on \emph{data} queries \textbf{(D)}, with a support of \emph{general joins} \textbf{(J)}. 
No special effort is shown to statistical queries and histograms, so they are considered unsupported.

\subsection{Differential Privacy based systems}
 Differential Privacy-based systems noisify the results of aggregation functions 
so as to make a person indistinguishable. The aim of a system is to calculate the noise 
as accurately as possible to keep the data usable. 
To achieve this, certain mechanisms redefine the notion of query sensitivity (Flex, Sample And Aggregate), 
and try to keep track of the transformations performed (PINQ, Qrlew). 
User-level systems (Qrlew) redefine the notion of distance to take into account 
all the contributions a person can make. 
A key feature of these systems is the ability to manage the privacy budget $\epsilon$ to respond to 
multiple requests, or to create more complex mechanisms through composition.

\subsubsection{PINQ}
PINQ~\cite{PINQ} is a framework for writing statistical queries \textbf{(A)} protected by \emph{DP} at \emph{tuple level}. 
A system using PINQ replaces the database by loading datasets and managing queries to access information via code 
offering SQL-like syntax, so PINQ has an \emph{integrated} architecture.

PINQ allows to perform transformations on data before it is protected by the DP mechanism. 
To ensure that the added noise remains adequate, PINQ relies on the notion of \emph{$c$-stable} transformation.
%
%

This definition is used to implement selection \textbf{(S)}, grouping by category
and the union of two relations.

It can also be used to implement 1-to-1 joins, \textit{i.e.}, joins where each tuple can only be associated with a unique other tuple. 
General joins do not have bounded stability because a tuple may be correlated with an unbounded number of other 
tuples, so a contribution may be taken into account an unknown number of times, so PINQ does not support them.
To solve this problem, wPINQ~\cite{wPINQ} adds a weight to each tuple to enable the impact of any join to be calculated. This system should therefore be studied in the future.

\subsubsection{Chorus}
Chorus \cite{JohnsonNHS20:Chorus} is a Scala library that provides tools to implement differential privacy for SQL queries with a \emph{proxy} architecture.
It enables the implementation of a wide range of mechanisms, by rewriting the SQL query or adding noise to the result, both parameterized by query analysis (e.g. sensitivity calculation). 

Chorus comes with some classic DP mechanisms.
This article will focus on two of them (SAA and \textsc{Flex}) usable as they stand 
(without development requiring a privacy expert). 
These mechanisms calculate the sensitivity of a query in a way other than \acrshort{GS}.
Mechanisms based on \acrshort{GS} cannot handle \emph{join}
\cite{PINQ} as they increase GS in an unbounded way, and others based on \acrshort{LS} cannot be differential private, because \acrshort{LS} can be very
different on two neighboring databases, so the noise level leaks information on the database~\cite{NissimRS07}. 
To resolve this the \acrfull{SS} \cite{NissimRS07} is an upper bound on \acrshort{LS} and is not related to a real
database, so the noise can be computed without leaking information on data.
However, computing \acrshort{SS} is NP-hard \cite{NissimRS07} for certain problems.
To solve this computability problem, these two mechanisms have different approaches: 
SAA proposes an algorithm in which SS is easy to compute,
\textsc{Flex} brings another definition\emph{elastic sensitivity} higher than SS but computable in linear time.

\paragraph{Sample And Aggregate:}

Sample And Aggregate (SAA) \cite{NissimRS07} is a framework that uses the definition of \acrlong{SS} to calibrate noise. 
As \acrshort{SS} is hard to compute, \acrshort{SAA} follows these steps: 
(i) partition the database (of size $n$) in $k$ sub-samples of size $n/k$;
(ii) apply the given statistical function of each partition; 
(iii) aggregate the result with a noisy function calibrated with \acrshort{SS}.

The aggregation function is chosen to make the computation of \acrshort{SS} easy.

\textsc{Chorus} implements \acrshort{SAA} by rewriting queries for step (i) and (ii) with a \verb_GROUP BY_ 
transformation, to create sub-samples.
This mechanism supports all aggregation functions \cite{NissimRS07} \textbf{(A)}, but does not support \emph{join} \cite{JohnsonNS18:Flex}. As a DP mechanism it can by definition protect \emph{selection} \textbf{(S)}.

\paragraph{\textsc{Flex}:}
\textsc{Flex} implements the \emph{elastic sensitivity} \cite{JohnsonNS18:Flex} mechanism. It is also an upper bound on \acrshort{LS} but not as tight as \acrshort{SS}. It can be
computed in linear time.
This sensitivity based mechanism supports general join \textbf{(J)} and selection \textbf{(S)} for statistical
queries \textbf{(A)}.
Natively, only \emph{count} aggregation is supported,
but \emph{sum} and \emph{average} can be implemented by adding new functionalities to the mechanism. 
The mechanism protects finite-domain histograms \textbf{(H1)} by retrieving all possible category key values directly from the database. 
Care must therefore be taken to ensure that no new values can be added to the database.

\subsubsection{Qrlew}
Qrlew~\cite{GrislainPS24:Qrlew} is a \emph{proxy} that rewrites SQL queries to provide \emph{differential privacy} at \emph{user-level}. 
The main idea of the mechanism is to represent the query by an execution graph, each node is rewritten considering 
their type. The graph is used to keep track of privacy units and propagates the ranges of attributes. This
is used to optimize noise.
The system uses the execution graph to find out whether there is a DP mechanism available at each moment of execution. 
If this is not possible, \textquote{Qrlew leverages synthetic data}~\cite{GrislainPS24:Qrlew}. 
This can be used for statistical queries \textbf{(A)} when DP data protection is not possible, or for data queries \textbf{(D)}.

Aggregation functions are rewritten in \emph{sum} functions protected by bounding contributions 
\cite{WilsonZLDSG20} and clipped to keep sensitivity bounded.
The system announces to implement \emph{"Bounded User Contribution"}\cite{WilsonZLDSG20}, so it should support 
\emph{general joins} \textbf{(J)} and protect infinite-domain histograms with \emph{$\tau$-thresholding} \textbf{(H2)}.

\subsubsection{PrivateSQL}
PrivateSQL~\cite{PrivateSQL}is presented as an integrated relational database protection system designed to provide a Differential Privacy SQL query execution engine.

Its main objective is to extend the application of DP to relational schemas. The system introduces a new generalization of DP for multi-relational data, allowing the granularity of the confidentiality unit to be defined at several resolutions (tuple-level or user-level). 
The methodological approach is based on the construction of private synopses from a set of selected views, which makes it possible to respond to an unlimited number of subsequent queries using only public synopses and not the private database, thereby minimizing cumulative confidentiality loss and improving the accuracy of results.

 For the types of queries supported, PrivateSQL is limited to count queries, but takes into account selections \textbf{(S)}, projections \textbf{(P)}, joins \textbf{(J)}, histograms \textbf{(H1 and H2)}, and correlated subqueries. 
The system can process all SQL structures thanks to its synopsis mechanism, which analysts can query freely.

\subsubsection{DOP-SQL}
DOP-SQL is a differentially private SQL query system designed to be general-purpose, extensible, and highly useful. It is based on the DP model with user-level protection. 
Unlike early DP SQL engines limited to simple aggregates, DOP-SQL supports all SPJA queries with optional group by, and no restrictions on the type of join. Aggregations covered include \verb+COUNT+, \verb+COUNT DISTINCT+, \verb+SUM+, \verb+MAX+/\verb+MIN+, and \verb+QUANTILE+.
The architecture is integrated with PostgreSQL, but its extensible design allows it to be adapted to other standard SQL engines. The system intercepts and rewrites queries using a parser that automatically completes the necessary joins with primary private relations, so that the privacy engine can apply the correct protection. 
The DP engine then dynamically chooses the optimal mechanism from a suite of recent algorithms, such as R2T~\cite{R2T} for sensitive joins, or Shifted Inverse~\cite{ShiftedInverse} for \verb+MAX+/\verb+MIN+/\verb+QUANTILE+.

\section{Qualitative Analysis}\label{sec:qualitative}

\begin{table*}[ht]
	\centering
	\begin{adjustbox}{width=0.85\textwidth}
	\begin{tabular}{ccccccccccc} 
		\toprule
		\textbf{System} & 
		\textbf{Model} & 
		\textbf{Unit} & 
		\textbf{Architecture} & 
		\multicolumn{6}{c}{\textbf{Queries types}} \\

		& & & 																															  & S	& P & J & A & D & H1 & H2 \\
		\midrule
		Diffix \cite{FrancisEM17:Diffix} & $k$-anonymity & User & Proxy  			  & x & - & - & x & - & -  & x  \\
		Open Diffix \cite{Diffix-Elm} & $k$-anonymity & User & Integrated 		  & x & x & - & x & - & -  & x \\
		PostgreSQL Anonymizer \cite{PostgreSQL_Anon}& $k$-anonymity & Tuple & Integrated 							& x & - & - & - & x & -  & -  \\
		PINQ \cite{PINQ} & DP & Tuple & Integrated 														& x &	- & 1-to-1 & x & - & -  & -  \\
		Chorus (\textsc{Flex} \cite{JohnsonNS18:Flex}) & DP & Tuple & Proxy		& x & - & x & x & - & x  & -  \\
		Chorus (\textsc{SAA} \cite{NissimRS07}) & DP & Tuple & Proxy   				& x & - & - & x & - & -  & -  \\
		Qrlew \cite{GrislainPS24:Qrlew} & DP & User & Proxy 									& x & - & x & x & x & -  & x  \\
		DOP-SQL~\cite{DOP-SQL} & DP & User & Integrated 									& x & x & x & x & - & -  & x  \\
		PrivateSQL~\cite{PrivateSQL} & DP & User & Integrated 									& x & x & x & x & - & x  & x  \\

		\bottomrule
	\end{tabular}
	\end{adjustbox}
		\caption{Classification of SQL Query Sanitization Systems by Technical Criteria}
		\label{table:summary}
\end{table*}

Table~\ref{table:summary} summarizes the classification of systems based on the criteria detailed in Section~\ref{sec:background}.
The analysis is structured along two dimensions: Model/Unit/Architecture and Query Support.
\begin{itemize}
    \item Model/Unit/Architecture: This columns compare the system's foundational elements: the protection model (k-anonymity or DP), the privacy unit (tuple or user), and the system architecture (proxy or integrated).
    \item Query Support: This details the specific relational operations and query types the system supports:
    \begin{itemize}
    \item \textbf{(S):} Indicates protection against singling out individuals through restrictive \emph{selection} predicates.
    \item \textbf{(P): } Confirms that mechanisms are appropriately parameterized by \emph{projected} attributes and optimized accordingly.
    \item \textbf{(J): } Denotes support for general joins, \verb_1-to-1_ marker means that the system only handles one-to-one joins, where each tuple is associated with at most one other tuple.
    \item \textbf{(A): } Indicates protection by adding noise to aggregation results.
    \item  \textbf{(D): }  Indicates protection  by modifying raw values (e.g., masking or generalization).
    \item \textbf{(H1/H2):} Specifies support for histograms over finite \textbf{(H1)} or non-finite/unknown \textbf{(H2)} category key domains.
\end{itemize}
\end{itemize}

Table~\ref{table:summary} demonstrates that none of the surveyed systems provides a universal solution for comprehensive privacy protection across all criteria.

Qrlew offers the greatest diversity in supported query types; however, its generalist approach—by transforming all aggregation functions into sums may pose efficiency trade-offs (as shown in next quantitative evaluation section).

Ultimately, the selection of an optimal system depends entirely on the specific analytical requirements and the constraints of the data environment. Key considerations include:
\begin{itemize}
    \item Analytical goal: Whether the analysis requires raw data \textbf{(D)} or only statistics \textbf{(A)}.

    \item Deployment feasibility: The practical possibility of using an \emph{integrated} architecture (requiring DBMS modification) versus a more flexible \emph{proxy}-based solution.

    \item Database structure: The complexity of the database (single-table vs. highly relational), which dictates the necessity and frequency of supporting general joins \textbf{(J)}.

    \item Privacy rigor: The choice between the weaker but often simpler $k$-anonymity model and the mathematically robust but more complex Differential Privacy model. Again with rigor in mind, there remain two questions.  
    Is the personal data stored at \emph{tuple-level} or \emph{user-level}? And if histograms are required, can the schema be used to define a finite domain \textbf{(H1)} or not \textbf{(H2)}?
\end{itemize}

Following the qualitative classification of sanitization systems, 
the analysis now shifts to a quantitative evaluation of their performance. 
This empirical study assesses the utility of systems on statistical queries involving aggregate functions like \verb+COUNT+ and \verb+SUM+. 
Consequently, the scope is narrowed to exclusively focus on systems based on Differential Privacy. The rationale for this selection is methodological: DP-based systems are designed to add calibrated noise to query results, which allows utility to be directly measured as the error between the sanitized and true values. 
In contrast, $k$-anonymity systems operate on raw data through generalization and suppression or simply reject queries that fail to meet a privacy threshold. 
Since these mechanisms do not produce a similarly measurable alteration of aggregate results, they are excluded from the subsequent performance analysis.

\section{Quantitative Analysis}\label{sec:quantitative}
This section assesses the performance and utility of a subset of the classified systems. We first detail the experimental methodology (Section~\ref{sub:methodology}), including the dataset, query set, and metrics used. We then present the results focusing on the trade-offs between privacy cost, query latency, and data accuracy (Section~\ref{sub:evaluation:utility}). Finally, we provide a critical analysis of the findings, linking empirical performance back to the theoretical criteria established in Section 2 (Section~\ref{sub:evaluation:overhead}).

\subsection{Evaluation Context}\label{sub:methodology}
Among the tools included in the qualitative study, only Chorus (Flex), Qrlew, and DOP-SQL were selected for experimentation.
The other tools are no longer maintained or are not suited to the needs of the experiment.

Diffix and OpenDiffix are no longer maintained, and the latest repository has compilation issues. 
Chorus (SAA) is a research prototype and has bugs that prevent it from being used.
PostgreSQL Anonymizer does not protect aggregation functions. The system works, but the experiment focuses on protecting aggregation functions, so it did not make sense to integrate this system.

PINQ's requirement to rewrite SQL queries into a \verb+C#+ 
syntax posed a significant integration challenge. While simple queries such as \verb+AVG+ and \verb+AVG+ were readily translated , more complex ones required either semantic modifications or substantial engineering effort. 
Consequently, supporting our use case for transparent SQL querying would have necessitated the development of a custom SQL-to-PINQ parser, which is out of scope of this work.

Finally, PrivateSQL was excluded from the experimental evaluation, as it lacks a publicly available implementation.

The systems are evaluated using a TPC-H database~\cite{TPC-H}, which provides a benchmark recognized within the database community.
This benchmark schema models a comprehensive sales and supply chain environment, comprising 8 distinct tables whose core entities include Customer, Order, Lineitem, Part, and Supplier. It incorporates a variety of data types, including numerical data (amounts, quantities), categorical attributes (status codes), and free-form text (descriptions, comments).

The TPC-H benchmark utilizes a standard suite of 22 complex analytical queries to rigorously evaluate the performance and scalability of DBMS. 
The structure of these queries necessitates sophisticated data processing, specifically incorporating intensive multi-way joins, complex data aggregations, and various conditional calculations.
Our data was generated with a scale factor of $1$, so the database contains approximately $8.10^6$ tuples.

TPC-H  provides queries ($Q_1$ - $Q_22$) among which some queries return raw personal data ($Q_2$, $Q_6$, $Q_9$, $Q_{12}$, $Q_{14}$, $Q_{15}$, $Q_{17}$, $Q_{19}$, $Q_{20}$, $Q_{22}$). DP systems cannot respond to non-statistical queries.
Among the remaining statistical queries, some use sub-queries ($Q_4$, $Q_7$, $Q_8$, $Q_{10}$, $Q_{11}$, $Q_{13}$, $Q_{16}$, $Q_{18}$, $Q_{21}$). However, the systems studied do not take this SQL structure into account.
Finally, among the few remaining queries, when there were multiple aggregation functions ($Q_1$) or grouping keys ($Q_3$), the systems were unable to respond. 
In the end, only query $Q_5$ can be answered, and only by DOP-SQL.

The queries that are not supported by any system are not described, but are included in the experiment repository.

Theoretically, queries that contain independent subqueries can be refactored by materializing their intermediate results. 
For instance, a scalar subquery can be pre-computed and its value substituted into the parent query. 
When a subquery returns a tabular result, it can be materialized as a temporary view or table for subsequent use. 
In practice, however, this approach may necessitate modifications to the existing database schema. 
It is crucial to note that this pre-computation technique is infeasible for correlated subqueries, as their evaluation is intrinsically dependent on row-level data from the outer query. 

However, our objective is to fairly evaluate these systems. 
This rewriting approach has thus not been implemented.

 To achieve the goal of a fair quantitative evaluation of these systems, we have provided a set of simple queries, which is divided into two parts: 
The former part is composed of “unit test” queries evaluating 
simple aggregation functions, generally on a single table to test  functionalities of the systems one by one. 
The latter part is composed of more complex queries and returns histograms, with aggregation functions, joins, and sub-queries. 
Table~\ref{table:queries} gives a short description for each evaluated query.

\begin{table*}
	\begin{tabular}{|c|l|l|}
		\hline
		\textbf{Test Type} & \textbf{Query} & \textbf{Description} \\
		\hline
		\multirow{6}{*}{{\textbf{Unit Test}}} & 
		 \verb_COUNT_ & Count of customers  \\
		\cline{2-3}
		& \verb|COUNT_DISTINCT| & Count of distinct nation with at least one order \\
		\cline{2-3}
		& \verb_SUM_ & Sum of customers's accounts   \\
        \cline{2-3}
		&\verb_SJA_ & Count of orders where the customer does not have sufficient funds in their account. \\
		\cline{2-3}
        & \verb_AVG_ & Average of the price of orders \\
		\cline{2-3}
		& \verb_MIN_ & Minimum cost of an order (\verb_orders.totalprice_) \\
		\hline
		\multirow{3}{*}{{\textbf{Histogram}}} & 
		\verb_histogram_ & Count of orders by status \\
		\cline{2-3}
		& \verb|ME_94_revenue| & Revenue by country in the middle east of the year 1994 ($Q_5$ of TPC-H) \\
		\cline{2-3}
		 & \verb|region_revenue| 
		& Revenue by region \\
		\hline
	\end{tabular}
    \caption{Specification of the Query Set Used for Experimental Evaluation}\label{table:queries}
\end{table*}

We conducted the experiments using a notebook with Python 3.11.2 (source code is accessible~\cite{DifPriPos_Sok}) on Linux Debian 12. 
The tests were performed on a machine equipped with an i9-13950HX processor with 32 cores 
clocked at 5.50 GHz and 64 GB of RAM.

\paragraph{Limitations}
The queries were tested in a “one query, one execution” mode. For some systems (PINQ, Chorus, Qrlew), it is possible with little development to execute queries in “online” mode with the system running continuously and receiving several queries one after the other.
For a system to be in online mode, it must handle the composition of mechanisms.
We did not perform this test because we wanted to execute each query under the same conditions (with a fixed budget and the possibility of using it entirely).
\par
Our experiments are based solely on statistical queries, not data queries. Knowing that only DP systems take statistics into account, we focused our tests on a DP approach.
\\

The following sections will focus on two types of evaluation. The first concerns the evaluation of the utility of sanitation mechanisms (Section~\ref{sub:evaluation:utility}). As such, we will measure the divergences between the original results and the results after their application. The second aims to measure the algorithmic overhead of such an implementation (Section~\ref{sub:evaluation:overhead}). This will be measured by means of the execution times of the different methods on a common platform.

\subsection{Utility Evaluation}\label{sub:evaluation:utility}

This section focuses on utility evaluation. 
It starts with presenting which metrics are used for scalar queries  (Section~\ref{subsub:scalar}), and  
for histogram queries (Section~\ref{subsub:histo}).
In both cases, $N=25$ repetitions of the evaluation of the probabilistic sanitization process are conducted. 
This repetition allows for the computation of the mean and the expression of the standard deviation (std) as a measure of dispersion.
Results are finally presented and analyzed (Section~\ref{subsub:results}).

\subsubsection{Scalar Statistical Queries}\label{subsub:scalar}

For a given scalar statistical query $Q$ (e.g., a sum calculation), the Relative Error ($RE$)
 is used as the evaluation metric. 
This error is defined as the absolute difference between the true (original) value, $F_Q$, and the corresponding sanitized value, 
$S_Q$, normalized by $F_Q$. 
Since the sanitization process is probabilistic, this evaluation is repeated $N$ times. 
The mean relative error ($MRE(Q)$) is then calculated and expressed as a percentage using the following formula:

\begin{equation}
MRE(Q) = \frac{100}{N} \sum^{N}_{i=1}{\frac{|S_Q^i - F_Q|}{F_Q}}
\end{equation}

\subsubsection{Histogram queries}\label{subsub:histo}

Consider a query $Q$ whose result is a histogram composed of $n$ frequencies, $H_1, \dots, H_n$, corresponding to $n$ distinct categories. The probabilistic sanitization process yields the sanitized frequencies $H'_1, \dots, H'_n$ for these categories. The quality of the sanitization for this histogram query is assessed using the {Mean Average Percentage Error (MAPE)}.

Since the sanitization is repeated $N$ times, the overall quality is measured by the mean of these MAPE values, denoted as the {MMAPE ($Q$)}. The formula for the MMAPE is thus defined as:

\begin{equation}
MMAPE(Q) = \frac{100}{N\times n} \sum_{i=1}^{N} \sum^{n}_{j=1} \frac{|H'^{i}_{j} - H_j|}{H_j} 
\end{equation}

\subsubsection{Results}\label{subsub:results}
\begin{figure*}[ht]
	\centering
	\begin{subfigure}{0.33\textwidth}
	\includegraphics[width=\textwidth]{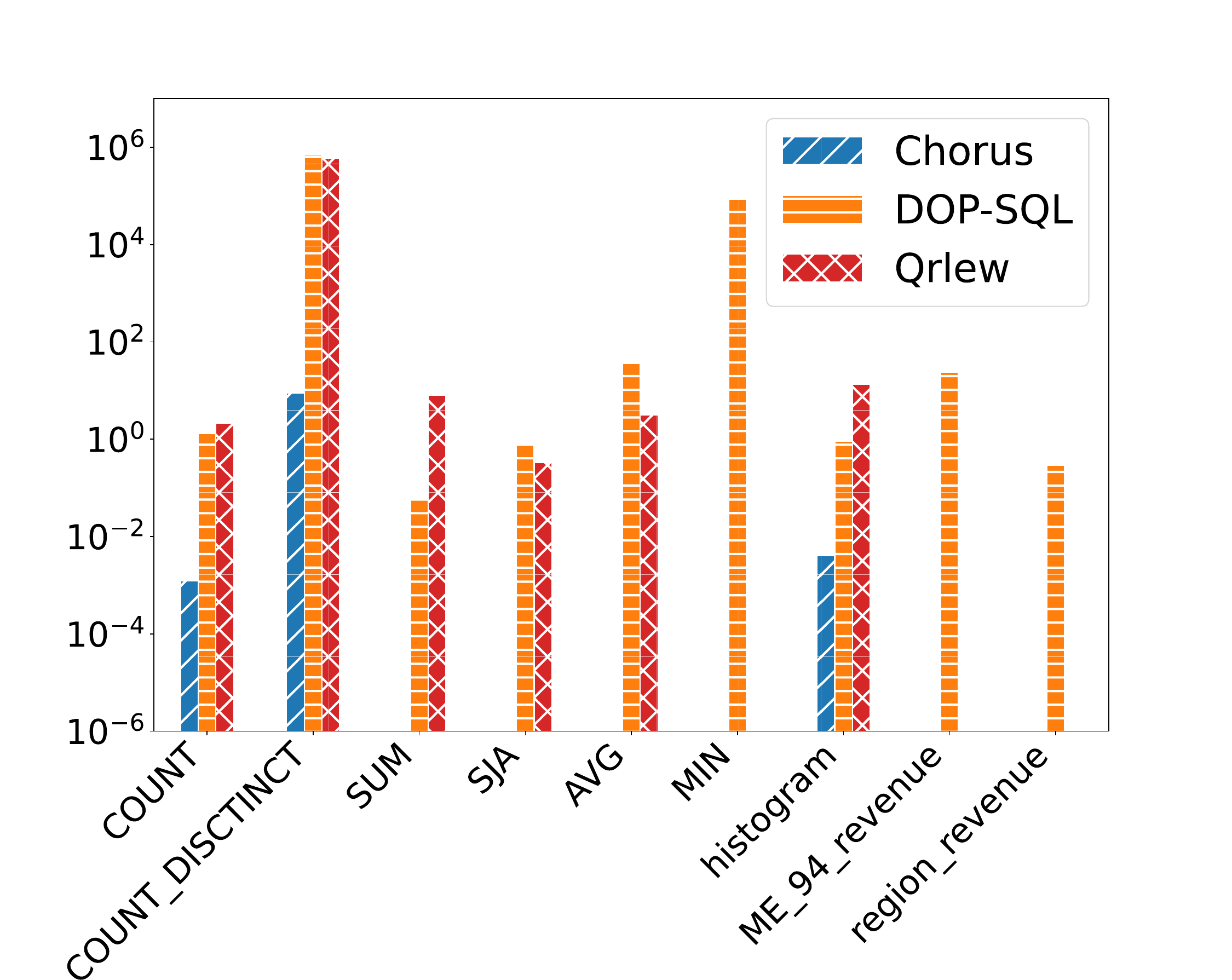}
		\caption{$\varepsilon = 0.1$}
	\end{subfigure}
	\begin{subfigure}{0.33\textwidth}
	\includegraphics[width=\textwidth]{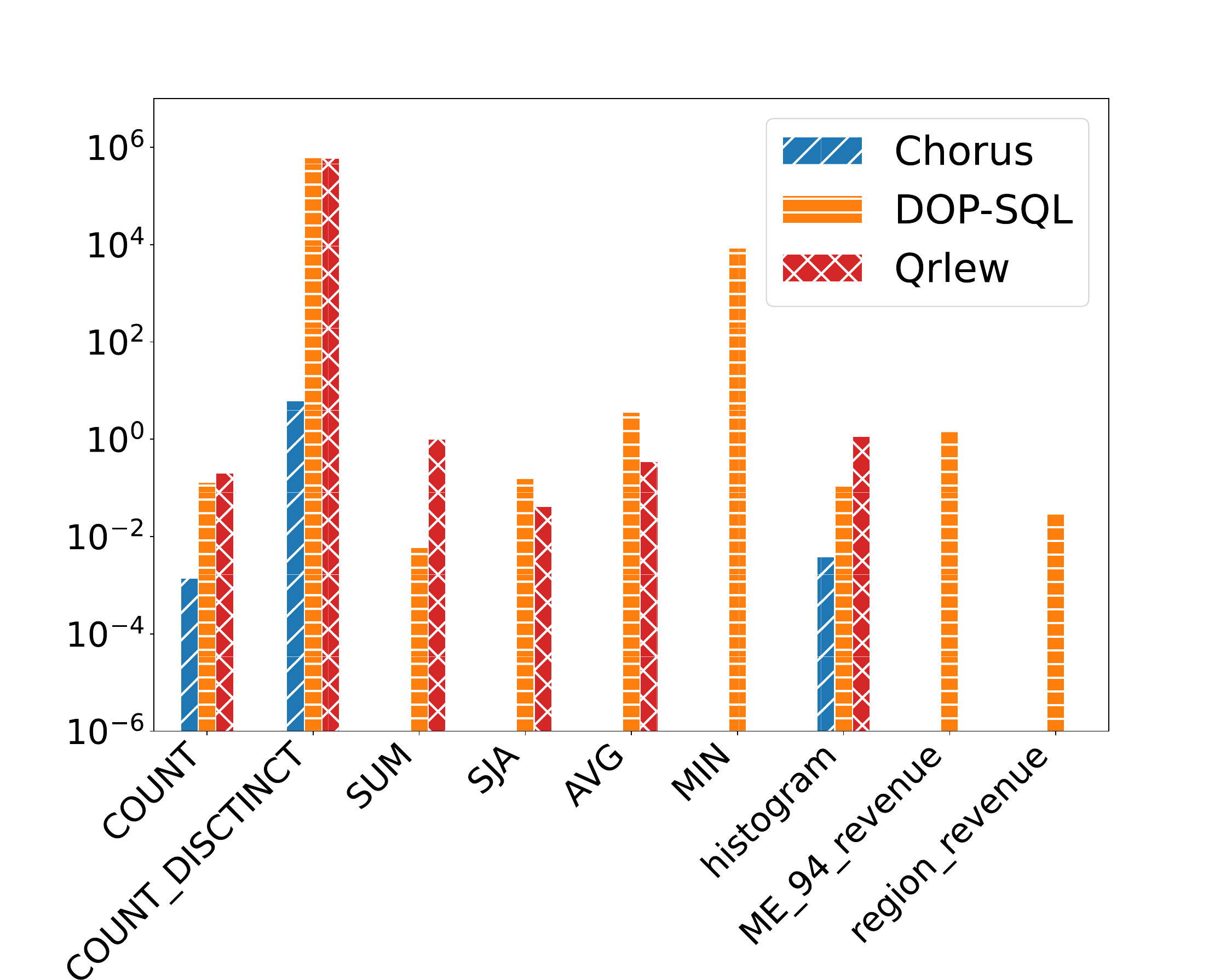}
		\caption{$\varepsilon = 1$}
	\end{subfigure}
	\begin{subfigure}{0.33\textwidth}
	\includegraphics[width=\textwidth]{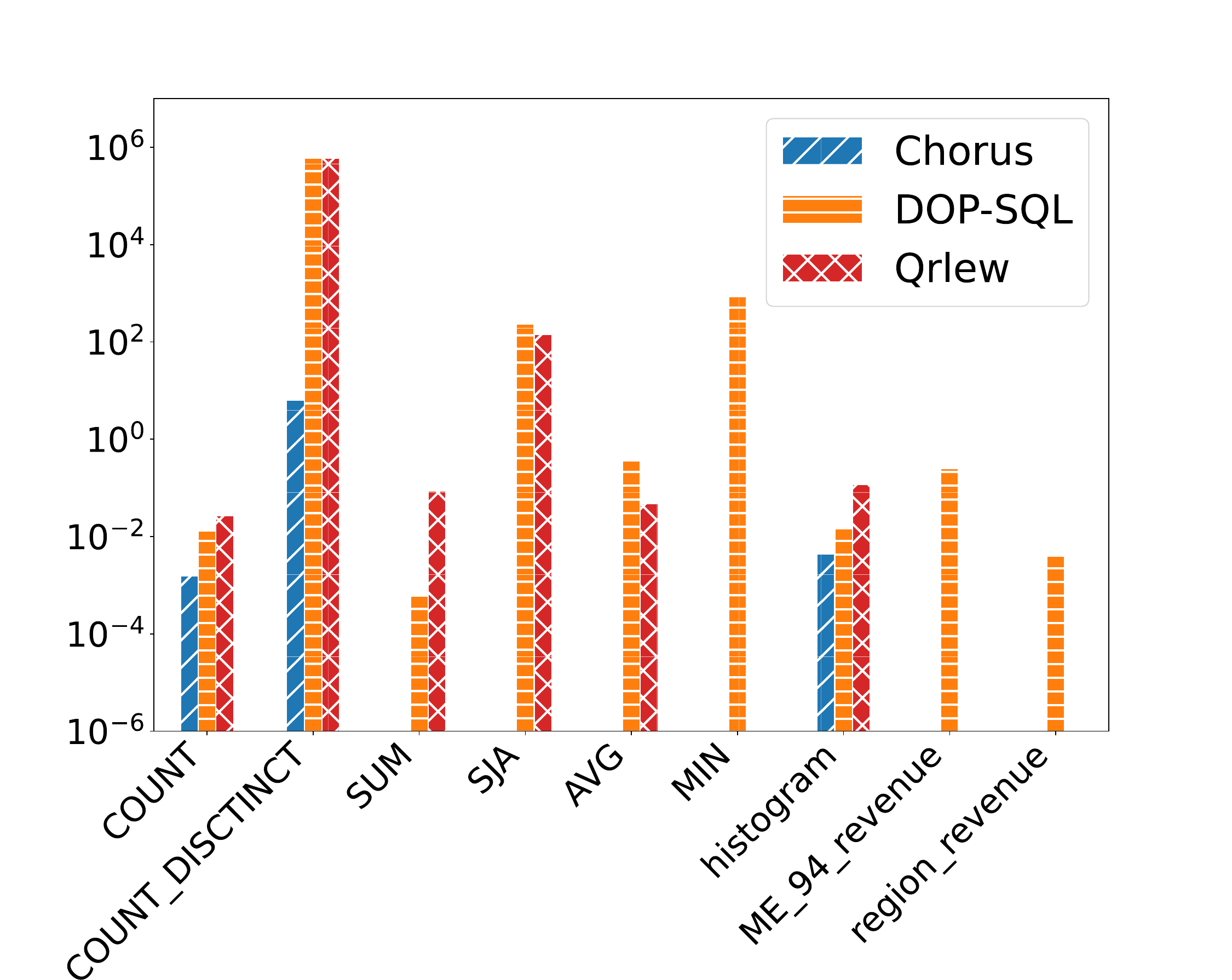}
		\caption{$\varepsilon = 10$}
	\end{subfigure}
    \caption{Percentage error for evaluated systems, comparing sanitized query results to original results as a function of the privacy budget $\varepsilon$. The y-axis uses a logarithmic scale.}

\label{fig:result:mape}

	\end{figure*}

Figure~\ref{fig:result:mape} presents the mean absolute percentage error (MAPE) for the systems.

The results show that the systems add more errors, and therefore usefulness decreases if the budget is lower. This is an expected result related to the definition of DP. 
Chorus produces results with less error, but can only respond to queries \verb|COUNT|, \verb|COUNT_DISTINCT| and \verb|histogram|, i.e., queries with only COUNT as aggregation function.
\\
Qrlew does not respond to all queries. This is due to its mechanism, where every aggregation function is rewritten with sums, which are then clipped and protected. So the protection mechanism is always the same.
This is why, for example, MIN and MAX are not supported, as they cannot be written with sums. 
\\
Unlike Qrlew, DOP-SQL chooses the mechanism to apply based on the query pattern, which explains why it provides better coverage of different query types. 
However, it is noteworthy that DOP-SQL performs better than Qrlew in terms of usefulness for sums, even though this function is at the heart of the Qrlew method.
\par
The systems significantly degrade the data on the SJA query. The join increases the sensitivity of the query in an unbounded manner. The systems therefore propose a bound on sensitivity and choose the noise accordingly, which is often very significant.
\\
DOP-SQL and Qrlew generate a lot of noise on the \verb|COUNT_DISTINCT| query. This is because these systems do not support the DISTINCT keyword, and therefore evaluate the COUNT with duplicates.
DOP-SQL and Qrlew should not respond to this query.
\par
It is important to note that systems sometimes have free parameters, which can also influence the results. We did not take them into account in this study because only the budget $\varepsilon$ and $\delta$ are common to all DP systems we have tested.

\subsection{Execution Time Overhead}\label{sub:evaluation:overhead}

\begin{figure}[ht]

	\centering
	\begin{subfigure}{0.40\textwidth}
	\includegraphics[width=\textwidth]{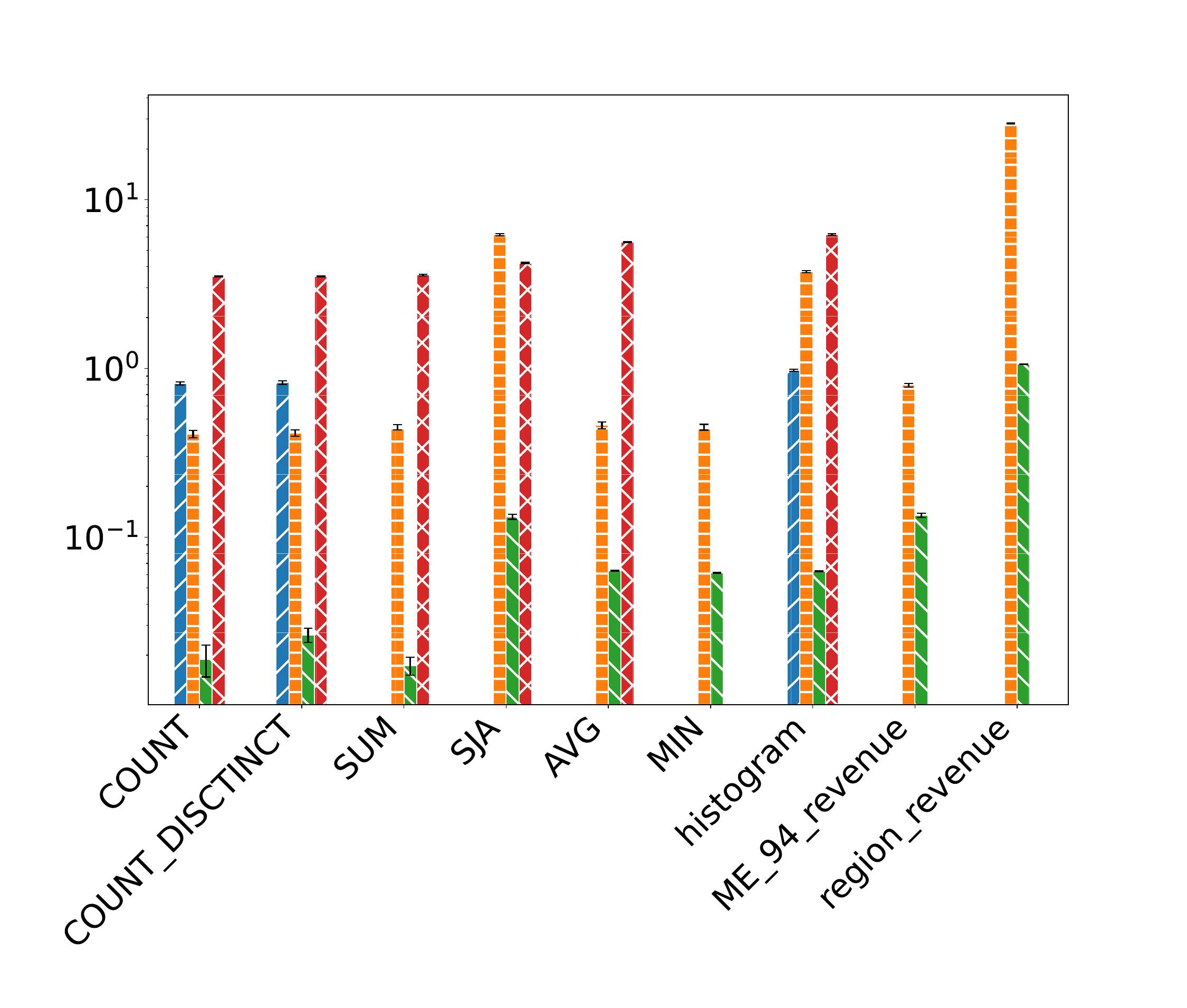}
	\caption{Execution Time (s)}
         \label{fig:res:time:brut}
	\end{subfigure}
	\begin{subfigure}{0.40\textwidth}
		\includegraphics[width=\textwidth]{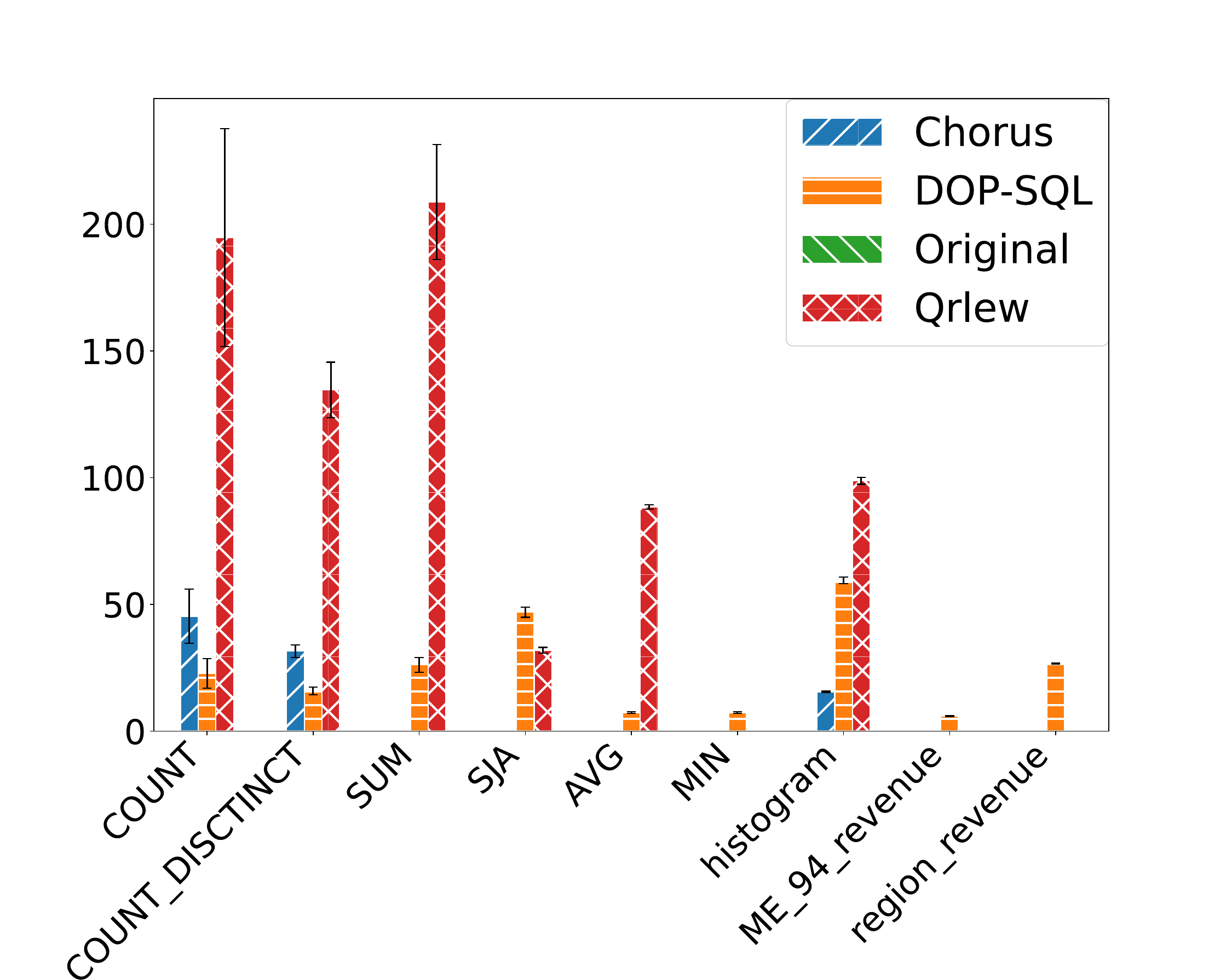}
	    \caption{Rate of Execution}
         \label{fig:res:time:rate}
	\end{subfigure}
	\caption{Execution time (s) and ratio of execution time overhead per system and per query. }\label{fig:result:time}
\end{figure}


This section analyzes the computational overhead introduced by the sanitization systems. 
For each query presented in the preceding section, execution times were recorded over $N=25$ repetitions to compute the mean and standard deviation (std). 
Figure~\ref{fig:result:time} synthesizes these results, presenting the absolute mean execution time (Figure~\ref{fig:res:time:brut}) and the overhead ratio, defined as the sanitized execution time divided by the original query time (Figure~\ref{fig:res:time:brut}).

The results reveal distinct performance profiles among the systems. 
Chorus and DOP-SQL exhibit efficient execution for simple Unit Test queries, with times on the order of one second. 
However, DOP-SQL's latency increases significantly for more complex queries involving joins and histograms. This behavior is attributed to its reliance on an embedded Linear Programming solver to dynamically optimize the selection of privacy mechanisms~\cite{R2T}.

In sharp contrast, Qrlew demonstrates a consistently high execution time, averaging around 10 seconds, across all query types, irrespective of their intrinsic complexity. This uniform performance stems directly from its core methodology: transforming all aggregation functions into a series of sum operations that are then sanitized by a DP mechanism. 
This standardized process results in a similar computational path and, consequently, similar execution times for all queries.

In general, systems are slower to execute queries because they calculate the exact value of the query and add noise, with a time-consuming calculation to best calibrate this noise. 
However, in the end, the result is less useful than the actual value, and therefore it takes longer to degrade a statistic that we had calculated accurately. 

Finally, it is noteworthy that the standard deviation values, illustrated by the error bars in Figure~\ref{fig:res:time:brut}, are consistently small relative to the mean execution times. This indicates the robustness and repeatability of the performance evaluation process.

\section{Discussion}\label{sec:discussion}
A qualitative analysis of query type coverage suggests that Qrlew, DOP-SQL, and PrivateSQL are comprehensive systems (see Table~\ref{table:summary}). However, empirical evaluation reveals significant practical limitations: Qrlew and DOP-SQL exhibit high execution latency and notable data degradation, while PrivateSQL lacks a public implementation, precluding its experimental assessment.

A significant limitation of the evaluated systems is their failure to support complex SQL structures. Specifically, no system could process queries containing either multiple aggregate functions or subqueries (both independent and correlated). This suggests that the systems utilize SQL merely as a high-level interface for selecting statistical computations and target tables, rather than supporting the language's full expressive power. Consequently, advanced applications of the SQL language are precluded by these protection systems.

Regarding performance, an increase in query execution time is an expected overhead, as privacy-preserving mechanisms inherently require additional computation compared to standard query execution. However, the efficiency of the prevailing method is questionable: these systems first compute an exact query result only to subsequently sanitize it.
Arguably, the performance overhead represents an acceptable trade-off for strong privacy guarantees, particularly for non-real-time analytical workloads. This argument is weakened, however, when the increased latency is coupled with a significant degradation in data utility, as was observed for certain queries with Qrlew and DOP-SQL.
While systems like PrivateSQL mention an alternative using approximate query calculation, the lack of a public implementation prevented us from empirically verifying whether this approach reduces computation time.

Ultimately, our results illustrate a recurring challenge in the deployment of differential privacy: the transition from elegant algorithms that function on simple queries (such as a COUNT or SUM) to practical, general-purpose systems that are genuinely usable by non-experts. The gap between the theoretical coverage of SPJA-type queries and their effective support in practice remains a major barrier to widespread adoption.

\section{Conclusion}\label{sec:conclusion}

This paper presents a qualitative and quantitative analysis of systems designed for sanitizing SQL queries. 
Our qualitative classification is based on a multi-criteria evaluation of each system, assessing foundational elements such as the privacy model and unit, the software architecture, and the breadth of supported SQL query types including selection, join, and aggregation.
Such a qualitative analysis demonstrates that no single system provides a universal solution for comprehensive privacy protection across all query types. 
A significant limitation observed across the board is the failure to support complex SQL structures, such as queries with multiple aggregate functions or subqueries. 

The subsequent empirical evaluation of selected systems reveals significant practical limitations. 
We identify a recurring trade-off where broader query support is often associated with high execution latency and notable data degradation. 
These results highlight the considerable gap between the theoretical aim of protecting complex SPJA-type queries and the effective support available in practice, which remains a major barrier to widespread adoption by non-experts.

Based on the results of our quantitative evaluation, we have identified the integration of common subexpression elimination into the query planner—specifically for handling independent subqueries—as a crucial enhancement. This development is expected to significantly expand the operational scope of the sanitization system and therefore constitutes our most immediate objective for future research.

A more ambitious research direction is to design a query engine where privacy protection is integrated directly into the planning and execution phases, following a privacy-by-design approach.  Such a system would address the potential inefficiency of the prevailing methodology, which first computes an exact result and then sanitizes it with noise. 
This could involve exploring approximate query processing (AQP) techniques to obviate the need for computing a precise result unnecessarily, thereby reducing latency while guaranteeing confidentiality.

\section{Artifacts}
All the source code for experiments can be found on SoftwareHeritage~\cite{DifPriPos_Sok}.
The experiments are as reproducible as possible, and all experimental results come from public code.

\bibliography{biblio.bib}
\end{document}